\documentclass[prd,nofootinbib,eqsecnum,preprintnumbers,11pt]{revtex4-1}

\usepackage{amssymb,amsmath,amsfonts,amsbsy,graphicx,bm}

\newcommand{\calN}{{\cal N}}
\newcommand{\calO}{{\cal O}}
\newcommand{\calR}{{\cal R}}
\newcommand{\calT}{{\cal T}}

\begin{document}
\

\preprint{\vtop{\hbox{CERN-PH-TH/2011-123}
\hbox{\hspace{17mm} YITP-11-57}}}

\title{Conformal invariance of curvature perturbation}

\author{Jinn-Ouk Gong$^\dag$\footnote{jinn-ouk.gong@cern.ch}}
\author{Jai-chan Hwang$^\ddag$\footnote{jchan@knu.ac.kr}}
\author{Wan Il Park$^\sharp$\footnote{wipark@kias.re.kr}}
\author{Misao Sasaki$^{\sharp,\S}$\footnote{misao@yukawa.kyoto-u.ac.jp}}
\author{Yong-Seon Song$^\sharp$\footnote{ysong@kias.re.kr}}

\affiliation{${}^\dag$ Theory Division, CERN,
CH-1211 Gen\`eve 23, Switzerland
\\
${}^\ddag$ Department of Astronomy and Atmospheric Sciences,
Kyungpook National University, Daegu 702-701, Republic of Korea
\\
${}^\sharp$ Korea Institute for Advanced Study,
Seoul 130-722, Republic of Korea
\\
${}^\S$ Yukawa Institute for Theoretical Physics,
Kyoto University, Kyoto 606-8502, Japan
}

\date{\today}

\begin{abstract}

We show that in the single component situation all perturbation
variables in the comoving gauge are conformally invariant to all
perturbation orders. Generally we identify a special time slicing, the
uniform-conformal transformation slicing, where all perturbations are
again conformally invariant to all perturbation orders. We apply
this result to the $\delta{N}$ formalism, and show its conformal
invariance.

\end{abstract}

\maketitle

\section{Introduction}

A large class of extensions of general relativity is described in the
context of scalar-tensor theories of gravity~\cite{STreview}, where a
space-time metric $g_{\mu\nu}$ is coupled to a scalar field $\phi$,
with the other matter contents such as fermion fields being minimally
coupled to gravity. By an appropriate conformal transformation
\begin{equation}\label{ct}
g_{\mu\nu} \to \overline{g}_{\mu\nu} = \Omega^2g_{\mu\nu} \, ,
\end{equation}
we can move to another frame. A prime example is the Einstein frame,
where after the conformal transformation (\ref{ct}) the gravitational
 action becomes the Einstein-Hilbert one. These conformal frames are
mathematically equivalent, so one can work in any frame as long as
mathematical manipulations are concerned. It is, however, not explicitly
clear whether these conformally related frames also enjoy physical
equivalence~\cite{Deruelle:2010ht}.

In cosmology, we encounter various frames of the metric which are
related by conformal transformation, such as Jordan frame,
Einstein frame, string frame and so on. The physical equivalence between
 these frames is especially important, since the cosmological observations
 with improving sensitivity can probe gravitational theories on the
largest observable scales~\cite{Song:2008vm}. Further, some models of
the early universe which accounts for the primordial perturbations
incorporate non-minimal coupling between gravity and a scalar field,
such as the recently proposed standard model Higgs
inflation~\cite{Bezrukov:2007ep}. Usually one moves from the original,
Jordan frame to the Einstein frame by a conformal transformation,
 and then computes perturbations there. The invariance of the vector
and tensor perturbations, and that of the scalar perturbation in the
comoving gauge, are shown up to second order~\cite{equivalence}, but
 it is not clear if this invariance holds fully non-perturbatively.
 Regarding precise upcoming experiments that may detect non-linear
signatures such as non-Gaussianity, it is important to clarify
the full non-linear invariance of the perturbations~\cite{Chiba:2008ia}.

In this article, we study the conformal invariance of the
cosmological perturbation. Our particular attention will be given
to the curvature perturbation on super-horizon scales.
We will show that in the single component case
all perturbations in the comoving slicing are conformally
invariant to fully non-linear order. Once this is given,
we can see that in the context of the $\delta{N}$ formalism
it is a matter of gauge transformation between
different coordinate systems which impose different slicing conditions.

As a temporal gauge condition, we take the conformal
transformation factor $\Omega$ to be unperturbed to
full non-linear order. This may be called as the
uniform-conformal-transformation slicing, or, uniform $\Omega$ slicing
(U$\Omega$S). It is convenient to decompose $\Omega$ into the
background and perturbation as $\Omega = \Omega_0e^\omega$, then
the U$\Omega$S means $\omega=0$ as the slicing condition. Thus we have
\begin{equation}
\Omega = \Omega_0(t) \, .
\end{equation}
Under this slicing condition it is obvious, almost a tautology, that
all perturbed quantities are naturally invariant under the
conformation transformation. This result applies
even in multiple component situation: see
(\ref{UOmegaG-fR}) and (\ref{UOmegaG-Fphi}) for relations implied
by this slicing condition to non-linear orders of perturbation.
Note that we may consider the action given by
\begin{equation}
   S=\int d^4x\sqrt{-g}\left[{1 \over 2} f (\phi,R)
-\frac{1}{2} \omega (\phi)g^{\mu\nu} \phi_{,\mu}\phi_{,\nu}-V(\phi)\right]\, .
\end{equation}
Here, $\Omega$ is a function of either the field $\phi$ for
$f=F(\phi) R$-type gravity or a function of $F$ or $R$ for $f(R)$-type
gravity, where
\begin{equation}
F\equiv \frac{\partial f}{\partial R}\,.
\label{defF}
\end{equation}
In these cases, the U$\Omega$S corresponds to the uniform-field
slicing or the uniform-$F$ slicing, respectively, and we
may call it as the UFS.

\section{single field case}
\label{sec:single}

First, we consider the single field $F(\phi)R$-type gravity in
which we can show the invariance, especially that of the curvature
perturbation.
As for our spatial metric, we write
\begin{equation}\label{spatialmetric}
\gamma_{ij} = a^2e^{2\calR}\widetilde{\gamma}_{ij} \, ,
\end{equation}
where $\widetilde{\gamma}_{ij}$ includes vector and tensor
perturbations. We choose $\widetilde\gamma_{ij}$ such that
it remains invariant under the conformal transformation
as described in detail in Appendix~\ref{app:confgamma}.
Then, under the conformal transformation (\ref{ct}),
(\ref{spatialmetric}) becomes
\begin{equation}\label{conformalspatialmetric}
\overline{\gamma}_{ij}
= \Omega^2a^2e^{2\calR}\widetilde{\gamma}_{ij}
 = a^2\Omega_0^2\exp\left[2(\calR+\omega)\right]\widetilde{\gamma}_{ij}
\equiv a^2\Omega_0^2\exp\left(2\overline{\calR}\right)\widetilde{\gamma}_{ij} \,,
\end{equation}
so that the spatial curvature perturbation in the conformally
transformed frame becomes simply
\begin{equation}
\overline{\calR} = \calR+\omega\,.
\label{ctcalR}
\end{equation}
 Thus, in the U$\Omega$S in which $\omega=0$,
we have $\overline{\calR} = \calR$. Further, as spelled out
in Appendix~\ref{app:confgamma}, the vector and tensor
perturbations in $\widetilde{\gamma}$ defined with respect to
a fixed fiducial background metric are also conformally
invariant to fully non-linear order.

In the single component case, the U$\Omega$S implies $\delta\phi=0$,
the uniform field gauge or the comoving gauge for $F(\phi)R$-type
gravity: in most scalar-tensor theories of gravity, the conformal
transformation is a function of $\phi$, $\Omega = \Omega(\phi)$.
This result has an important implication that, for $F(\phi)R$
gravity~\cite{Hwang:2001qk} all perturbation variables in the
uniform field gauge are conformally invariant. In
(\ref{conformalspatialmetric}), we have explicitly shown the
invariance to all perturbation orders, and also have shown that why
the invariance is natural in the U$\Omega$S which is the same as the
uniform field gauge, or equivalently comoving gauge,
in the single component case.

Now, we provide a complementary point of view using the $\delta{N}$
 formalism~\cite{Starobinsky:1986fxa,Sasaki:1995aw,Lyth:2004gb}
in the universe dominated by a single scalar
field $\phi$. 
Since we will be interested in super-horizon scales in terms of
the $\delta{N}$ formalism, we restrict ourselves to zeroth order
in $\epsilon \equiv k/(aH)$, where $k$ is the comoving wavenumber
of a characteristic perturbation of our interest.
We consider a point on a comoving slice on which
$\delta\phi=0$. Hereafter we use the subscript $c$ to denote the
quantities evaluated on the comoving slices, and likewise $f$ and
$\overline{f}$ on the flat and $\overline{\text{flat}}$ slices (see
below). When the universe is dominated by a single
component $\phi$, the comoving slice is the same as the
U$\Omega$S, thus $\overline{\calR}_c = \calR_c$. That
is, the comoving curvature perturbation is the same and is thus
conformally invariant.

Until this point we have not yet related the perturbation in the
number of $e$-folds $\delta{N}$ to $\calR_c$. We can compute $N$ in
the standard manner. From (\ref{spatialmetric}), the expansion
scalar $\theta = 3H$, viz. the ``local'' Hubble parameter is given
by
\begin{equation}
H = \frac{1}{\calN}\frac{d}{dt}\log\left(ae^\calR\right) \, ,
\end{equation}
where $\calN$ is the lapse function.
From this, the number of $e$-folds $N$ is given by the integral
of $H$ along the proper time $d\tau=\calN dt$ as
\begin{equation}
N = \int_i^e Hd\tau = \log \left[ \frac{a(t_e)}{a(t_i)} \right]
 + \calR(t_e) - \calR(t_i) \equiv N_0 + \calR(t_e) - \calR(t_i) \, .
\end{equation}
Here, $t_i$ and $t_e$ denote the initial and the final moments respectively.

Now, we consider the conformally transformed frame. We can take
similar steps to obtain
 $\overline{H} = d\log\left(a\Omega e^\calR\right)/(\calN\Omega dt)$
and $d\overline{\tau} = \calN\Omega dt$, so that the number of $e$-folds
 in this frame $\overline{N}$ becomes
\begin{equation}
\overline{N} = \int_i^e \overline{H}d\overline{\tau}
= \log \left[ \frac{(a\Omega_0)(t_e)}{(a\Omega_0)(t_i)} \right]
 + \overline{\calR}(t_e) - \overline{\calR}(t_i) \, .
\end{equation}
Note that the background is {\em different} from $N_0$ as
\begin{equation}
\overline{N}_0 \equiv
\log \left[ \frac{(a\Omega_0)(t_e)}{(a\Omega_0)(t_i)} \right]
 = N_0 + \log \left[ \frac{\Omega_0(t_e)}{\Omega_0(t_i)} \right]
 \equiv N_0 + \Delta_0 \, .
\end{equation}

Here, we make a specific choice for the initial and final moments
where we evaluate the number of $e$-folds and the curvature
perturbation. As a reference point, we set the final moments
in both frames to be identical on a comoving slice. That is,
$\phi_c(t=t_e) = \phi_0(t_e)$. Thus, we have
$\omega(t_e) = \omega_c(t_e) = 0$ and in both frames
$\calR(t_e) = \overline{\calR}(t_e) = \calR_c(t_e)$.
This is the conformal invariance of $\calR_c$ we have seen above.
 Meanwhile, in the frame without overbar,
 we set the initial moment to be flat so that
$\calR(t_i) = \calR_f(t_i) = 0$, and in the other frame
``$\overline{\text{flat}}$'' in the sense that
$\overline{\calR}(t_{\overline{f}})
 = \overline{\calR}_f(t_{\overline{f}})
= \left( \calR+\omega \right)_{\overline{f}}(t_{\overline{f}})=0$.
Then, in the frame without overbar, we have
\begin{equation}\label{deltaNinv1}
N-N_0 \equiv \delta{N} = \calR_c(t_e) \, ,
\end{equation}
and in the frame with overbar,
\begin{equation}\label{deltaNinv2}
\overline{N}-\overline{N}_0 \equiv \delta\overline{N} = \calR_c(t_e) \, .
\end{equation}
That is, in both frames, $\delta{N} = \delta{\overline{N}} = \calR_c$.
Despite of the different background numbers of $e$-folds,
 their perturbations are invariant in both frames.

Notice that the curvature perturbation on a comoving slice $\calR_c$
and the field fluctuation on a flat slice $\delta\phi_f$ are related
by [see Eq.\ (287) in Ref.~\cite{Noh:2004bc}]
\begin{equation}\label{Rc_deltaphif}
-\calR_c
= \frac{H}{\dot\phi_0}\delta\phi_f
- \frac{H}{\dot\phi_0^2}\delta\phi_f\delta\dot\phi_f
+ \frac{H^2}{2\dot\phi_0^3}\frac{d}{dt}
\left( \frac{\dot\phi_0}{H} \right) \delta\phi_f^2 + \cdots \, .
\end{equation}
Here we remind that we are working at zeroth order in $k/(aH)$.
This relation holds at any arbitrary time. Meanwhile,
in the $\delta{N}$ formalism, by construction there is no field
fluctuation on the final comoving slice where we evaluate the
 curvature perturbation. Thus, we are to relate the curvature
perturbation on the {\em final} comoving slice $\calR_c(t_e)$ to
the field fluctuations on the {\em initial} flat slice
$\delta\phi_f(t_i)$. Furthermore, using the fact that we are interested
in the dynamics along a given background trajectory, we can show that
 up to second order in $\delta\phi$,
\begin{equation}\label{deltaN_deltaphideltaphidot}
-\delta{N} = \frac{H}{\dot\phi_0}\delta\phi
 - \frac{H}{\dot\phi_0^2}\delta\phi\delta\dot\phi
 + \frac{H^2}{2\dot\phi_0^3}
\frac{d}{dt}\left( \frac{\dot\phi_0}{H} \right) \delta\phi^2 + \cdots \, .
\end{equation}
We give detailed steps to find this relation in Appendix~\ref{app:phasespace}.
Once evaluated on a flat slice, this should be identical to $\calR_c(t_e)$
as we have shown in (\ref{deltaNinv1}) [or (\ref{deltaNinv2}) as well] and
 is in perfect agreement with (\ref{Rc_deltaphif}).
Thus, we conclude that $\calR_c(t_e) = \calR_c(t_i)$, i.e.
the curvature perturbation is conserved on large
scales\footnote{This could be regarded as an alternative proof
of the conservation of the non-linear curvature perturbation
as shown in Ref.~\cite{Lyth:2004gb} for a perfect fluid,
and in Ref.~\cite{Naruko:2011zk} for a generic scalar field.
However there is an important difference that here the separate universe
approach is assumed from the beginning, while its validity is explicitly
shown in Refs.~\cite{Lyth:2004gb,Naruko:2011zk}.}.

\begin{figure}[t]
 \begin{center}
  \includegraphics[width=12cm]{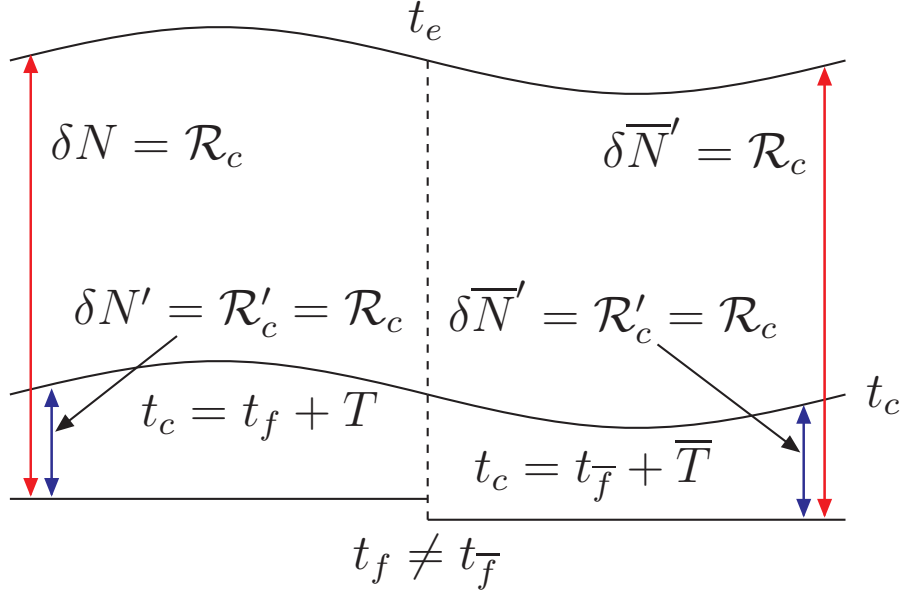}
 \vspace{-2em}
 \end{center}
 \caption{A schematic figure showing the different slices which are
connected by gauge transformations. Since $\calR_c$ is conserved on
large scales, $\calR_c(t_e) = \calR_c(t_c)$. The comoving slice on
which time is denoted by $t_c$ is common to both frames. 
Their initial slices are set in such a way that
in one frame we set the initial slice to satisfy $\calR(t_f) = 0$.
 In the other frame, on the other hand, we demand
$\overline{\calR}(t_{\overline{f}})
 = \left( \calR + \omega \right)(t_{\overline{f}}) = 0$.
We call the former and the latter to be ``flat'' and
 ``$\overline{\text{flat}}$'', and the quantities evaluated on these
 slices by a subscript $f$ and $\overline{f}$ respectively.}
 \label{fig:equivalence}
\end{figure}

Given the fully non-perturbative invariance of $\calR_c$ in two frames,
 we can obtain a useful perturbative relation as follows. For this,
it is convenient to insert another comoving slice common to the both
 frames between the initial and the final moments, and we denote by
 $t_c$ the time on this slice. Since we are interested in the large
scale limit where $\calR_c$ is conserved, the perturbations in the
number of $e$-folds between the initial moments and $t_c$ in two
frames, $\delta{N}'$ and $\delta{\overline{N}}'$, are identical
 to $\calR_c$. This is schematically shown in Fig.~\ref{fig:equivalence}.

Then, from the frame without overbar, suppressing the dependence
 of $N$ and $N_0$ on $\phi_0(t_c)$ which is common, for arbitrary $t<t_c$
 we have\footnote{Here, we have implicitly made a stronger
 assumption than (\ref{deltaN_deltaphideltaphidot}) that
$\dot\phi_0$ can be written as a function of $\phi_0$,
so that we have a single degree of freedom in the phase space.}
\begin{equation}
\calR_c(t_c) = N[\phi_f(t)] - N_0[\phi_0(t)]
 = \sum_n \frac{N_0^{(n)}[\phi_0(t)]}{n!} [\delta\phi_f(t)]^n \, ,
\end{equation}
where the superscript $(n)$ denotes $n$-th derivative and the
right hand side is evaluated on the flat slice.
Likewise, we can write another relation,
\begin{equation}
\calR_c(t_c)
= \sum_n \frac{N_0^{(n)}[\phi_0(t)]
+\Delta_0^{(n)}[\phi_0(t)]}{n!} \left[ \overline{\delta\phi}_f(t) \right]^n \, .
\end{equation}
Note that we have the same background field $\phi_0$, since it can
be redefined in one frame to absorb the background conformal
transformation factor $\Omega_0$. Then, essentially now we have
three different gauges conditions: $\calR+\omega=0$, $\calR = 0$
and $\delta\phi=0$,
which correspond to $\overline{\text{flat}}$, flat and comoving gauges respectively.
We need a specific conformal transformation
$\Omega(\phi)$ only to define the first gauge. Thus, using
appropriate coordinate transformation laws, we can freely move
between different gauges and relate one to another. For example,
we consider the coordinate transformation from $t_f$ and
$t_{\overline{f}}$ to $t_c$,
\begin{equation}\label{coordxform}
t_c = t_f + T(t_f,{\bm x})
 = t_{\overline{f}} + \overline{T}(t_{\overline{f}},{\bm x}) \, .
\end{equation}
Then, we only have to write the field fluctuations in terms of the
translation $T$ or $\overline{T}$. Since on the comoving slice there
 is no field fluctuation by construction, we have
\begin{equation}\label{phitc_phitf}
\phi_0(t_c) = \phi_0[t_f+T(t_f,{\bm x})]
\equiv \phi_f(t_f,{\bm x}) = \phi_0(t_f) + \delta\phi_f(t_f,{\bm x}) \, ,
\end{equation}
and a similar relation holds for $t_{\overline{f}}$.
Then, using (\ref{coordxform}) we can write
\begin{equation}
t_{\overline{f}} = t_f + \left[ T(t_f,{\bm x})
 - \overline{T}(t_{\overline{f}},{\bm x}) \right]
 \equiv t_f + \calT(t_f,{\bm x}) \, ,
\end{equation}
where $t_{\overline{f}}$ dependence of $\overline{T}$ on the right-hand
side can be removed by iteratively using this relation.
Then, from (\ref{phitc_phitf}), we can find the linear relation
between $\overline{\delta\phi}(t_f,{\bm x})$ and $\delta\phi(t_f,{\bm x})$ as
\begin{equation}
\overline{\delta\phi}_f(t_f,{\bm x})
 = \delta\phi_f(t_f,{\bm x}) - \dot\phi_0(t_f,{\bm x})\calT(t_f,{\bm x})
 + \cdots \, .
\end{equation}
This linear relation is in agreement with the well-known linear gauge
 transformation law~\cite{Kodama:1985bj} as it should be.
Note that in fact it does not matter that we necessarily restrict
ourselves to the transformation between two flat gauges,
 and hence we may drop all the subscript $f$. General coordinate
 transformations at arbitrary time would work as well.

\section{multi-field case}
\label{sec:multi}

In this section, we consider the multiple component situation.
We consider an action,
\begin{equation}
   S = \int dx^4 \sqrt{-g} \left[
       {1 \over 2} f\left(\phi^K, R\right)
       - {1 \over 2} G_{IJ} \left(\phi^K\right) g^{\mu\nu} \phi^I_{;\mu} \phi^J_{;\nu}
       - V\left(\phi^K\right) \right] \, .
   \label{L-original}
\end{equation}
Under a conformal transformation (\ref{ct}) with the field
redefinition $\Omega^2 \equiv F \equiv e^{\sqrt{2/3} \psi}$,
(\ref{L-original}) becomes
\begin{equation}
   S
       = \int d^4x \sqrt{- \overline{g}} \left[
       {1 \over 2} \overline{R}
       - {1 \over 2} \overline{g}^{\mu\nu} \psi_{,\mu} \psi_{,\nu}
       - {1 \over 2 F} G_{IJ} \overline{g}^{\mu\nu} \phi^I_{,\mu} \phi^J_{,\nu}
       - \overline{V} \right] \, ,
\end{equation}
where
\begin{equation}
   \overline{V} \equiv {1 \over 4 F^2} \left( 2 V - f + RF \right),
\end{equation}
with $F \equiv {\partial f/\partial R}$.
We have $\psi =\psi \left(\phi^K, R\right)$ in general.
We consider two cases:

\noindent
(I) The case with $f = f(R)$. In this case we have $\psi = \psi(R)$,
which is an additional scalar field minimally coupled to gravity
in the Einstein frame. An equivalent formulation in terms of
a Brans-Dicke type scalar field $\Phi$ is given in Appendix~\ref{sec:fR},
with its relation  to the canonical scalar field $\psi$ in the Einstein frame.

\noindent
(II) The case with $f = F\left(\phi^K\right) R$. In this case
we have $\psi = \psi\left(\phi^K\right)$. Hence there is no additional
scalar field. We have
\begin{equation}
   S = \int d^4x \sqrt{-\overline{g}} \left[
       {1 \over 2} \overline{R}
       - {1 \over 2} \overline{G}_{IJ} \overline{g}^{\mu\nu} \phi^I_{,\mu} \phi^J_{,\nu}
       - \overline{V} \right] \, ,
\end{equation}
where
\begin{equation}
   \overline{G}_{IJ}
       \equiv {1 \over F} G_{IJ} + \psi_{,I} \psi_{,J}.
\end{equation}
We see that the effect of the conformal transformation
is to change the non-linear sigma coupling term and the potential.
These conformal transformation properties are presented in
Appendix~A of Ref.~\cite{Hwang:2001qk}.

When the universe is dominated by a single component,
we have shown the conformal invariance of all perturbations
in the comoving slicing, which is equivalent to the U$\Omega$S.
In the multi-component situation
the U$\Omega$S is different from the uniform field slicing
$\delta\phi^I = 0$ of a chosen field component $I$
or the collective comoving slicing which sets $T^0{}_i=0$
in the energy-momentum tensor (or $u_i = 0$ in the
collective fluid four-vector. It should be noted however that
it is in general impossible to uniquely define the comoving slicing in
the multi-component situation because $u_{[\mu;\nu]}$ may not be zero). Consequently, the curvature
perturbation in the gauge where $\delta\phi^I=0$ is not
conformally invariant: denoting the gauge $\delta\phi^I=0$
 by a subindex $I$, to linear order, we have
\begin{equation}
\overline{\calR}_{I} = \calR_{I} + \sum_{J\neq I}
 \frac{\Omega_{,J}\delta\phi^J_{I}}{\Omega} \, ,
\end{equation}
where
\begin{equation}
\delta\phi^J_{I} = \delta\phi^J -
\dot\phi_0^J\frac{\delta\phi^I}{\dot\phi_0^I} =
-\frac{\dot\phi_0^J}{H}\left(\calR_{J} -\calR_{I}\right) \, ,
\end{equation}
so that clearly $\overline{\calR}_{I} \neq \calR_{I}$.

Nevertheless, even in the multiple component situation we can take
U$\Omega$S with $\omega \equiv 0$ to arbitrary non-linear order.
Under this slicing condition we have $\overline{\cal R} = {\cal R}$,
and similarly for all the other metric perturbation variables~\cite{Gong:2011cd}.
Let us denote quantities in U$\Omega$S by suffix $\omega$,
e.g. $\calR_\omega$.

For definiteness, let us consider the meaning of U$\Omega$S
in the two cases of our interest. Since $\omega \equiv 0$
implies $\delta F = 0 = \delta \psi$, we have the following:

\noindent
(I) For $F = F(R)$, we have
\begin{equation}
   \delta F = F_{,R} \delta R + {1 \over 2!} F_{,RR} \delta R^2 +
       \dots = 0 \, ,
   \label{UOmegaG-fR}
\end{equation}
thus $\delta R = 0$.

\noindent
(II) For $F = F\left(\phi^K\right)$ we have
\begin{equation}
   \delta F
       = F_{,I} \delta \phi^I
       + {1 \over 2!} F_{,IJ} \delta \phi^I \delta \phi^J
       + \dots
       = 0 \, .
   \label{UOmegaG-Fphi}
\end{equation}
This condition should be imposed among all the fields that appear
in $F$.

Here we note that, in the spirit of the $\delta N$ formalism,
the final comoving hypersurface should be chosen such that there
remains only a single adiabatic mode in the perturbation at and
after that epoch. Once the universe has entered this stage,
we also have the conservation of the comoving curvature perturbation
$\calR_c$. At this stage, the only dynamical degree
of freedom coupled to gravity should be encoded in the function $F$,
where the U$\Omega$S coincides with the comoving slicing (as discussed
below) on super-horizon scales, and hence
$\calR_\omega=\calR_c=$constant\footnote{Here we assume that $F$
and consequently $\Omega$ depend non-trivially on the remaining
dynamical degree of freedom. If it is not the case, then $\Omega$
will be simply a constant, and the conformal invariance is trivial.}.
All the other degrees of freedom should either be died out
by then or be purely isocurvature at and after that epoch in the sense
that they have no coupling to gravity whatsoever.

Conversely, at a stage during which multiple components of matter or
fields have their individual dynamics, $\calR_\omega$ is not conserved
yet. At this stage, whether some quantities are conformally invariant or
not is not really important. It is similar to the matter of
the gauge choice. Some quantities may be conserved in a certain gauge,
which may be a mathematically useful fact but has no physical/observational
significance. In the present case, $\calR_\omega$ and all the other
quantities defined in U$\Omega$S are conformally invariant. But
this would not help us much to understand the physics. It becomes
useful and meaningful only after $\calR_\omega$ becomes to be conserved.

Now we address the relation between the comoving slicing
condition and the U$\Omega$S in the case of $f(R)$ gravity,
\begin{equation}\label{eq:fR}
S=\int d^4x\sqrt{-g}\frac{1}{2}f(R)\,.
\end{equation}
In the case of $f(R)$-gravity the
comoving slicing {\it differs} from the UFS
in general even at linear order.
In the case of a minimally coupled scalar field we
have $T^0{}_i = \phi^{,0}\phi_{,i}$, thus $\delta \phi \equiv 0$
implies $T^0{}_i = 0$ to fully non-linear order which
is the comoving slicing condition. Whereas in the case of
pure $f(R)$ gravity, we have [see Eq.\ (3) in Ref.~\cite{Hwang:2001qk}]
\begin{equation}
   G_{\mu\nu} = {1 \over F} \left[ {1 \over 2} \left( f - R F
       \right) g_{\mu\nu}
       + F_{;\mu\nu} - g_{\mu\nu} F^{;\sigma}{}_{;\sigma} \right] \, .
\label{fReq}
\end{equation}
If we define the right hand side as an effective $T_{\mu\nu}$, to
the linear order we have [see Eqs.\ (18) and (B2) in
Ref.~\cite{Hwang:2001qk}]
\begin{equation}
   T^0{}_i = - {1 \over a F} \left( \delta \dot F - H \delta F -
       \dot F \alpha \right)_{,i} \, ,
\label{linT0i}
\end{equation}
where $g_{00} \equiv - a^2 ( 1 + 2 \alpha )$.
Thus, the UFS condition $\delta F = 0$ does not imply $T^0{}_i =0$
 in general in the original frame.

However, one can show that the UFS condition coincides with
the comoving slicing condition $T^0{}_i =0$ on super-horizon scales
to full non-linear order provided that the decaying mode becomes
negligible and there remains only a single adiabatic mode.
One can show this with the help of the field equation
for $F$ given by the trace of (\ref{fReq}).
One finds that it is identical to the background field equation
on super-horizon scales, i.e. at each spatial point ${\bm x}$
where a point means a Hubble size region, if one replaces
$t$ by $\tau$. Thus the general solution to full non-linear order
is given by $F=F\left(\tau(t,{\bm x}),{\bm x}\right)$. Then after the
decaying mode has disappeared, $\partial F/\partial\tau$
becomes a function of $F$ itself similar to the case of
standard slow-roll inflation. This implies that on UFS on which
$F=F_0(t)$, $\partial F/\partial\tau$ is also a function of
only $t$.
Then on UFS we have
\begin{equation}
T^0{}_i\propto n^\mu F_{;\mu\nu}P^\nu{}_{i}
=(F_{;\mu}n^\mu)_{;\nu}P^\nu{}_i
=\frac{\partial}{\partial x^i}
\left(\frac{\partial F}{\partial\tau}\right)=0\,,
\label{NLT0i}
\end{equation}
where $n^\mu$ is the unit normal to the uniform $F$ slice,
$P^\nu{}_\mu=\delta^\nu{}_\mu+n_\mu n^\nu$ is the
induced metric on the uniform $F$ slice, and
$\partial/\partial\tau=\partial/({\cal N}\partial t)$ is
the proper time along $x^i=$constant [the shift vector is
 $\calO(\epsilon)$ on super-horizon scales].
In fact, this leads to the conservation of the curvature
perturbation on the comoving slices~\cite{Naruko:2011zk}.

It may be instructive to spell out how this happens at linear order.
In the UFS the perturbed field equation gives~\cite{Hwang:2001qk}
\begin{equation}
   \dot {\cal R} = \left( H + {\dot F \over 2 F} \right) \alpha\,.
\end{equation}
Thus the fact that there remains only a single adiabatic mode
implies $\alpha=\calO(\epsilon^2)$, and hence $\dot {\cal R} = 0$
at leading order on super-horizon scales. Also the fact
$\alpha=\calO(\epsilon^2)$ implies $\tau=t$ at leading order.
Thus the field $F$ in UFS is actually given by
$F\left(\tau(t,{\bm x}),{\bm x}\right)=F(t,0)=F_0(t)$.

In the above, we have considered the case when the field $F$ dominates
the universe. In general, we may have other fields or matter which
may have non-trivial background dynamics. In this multi-component
situation, we have no conformal invariance of the comoving
curvature perturbation (if it can ever be sensibly defined).
In this situation, we can still define a conformally invariant
curvature perturbation, $\calR_\omega$ in U$\Omega$S.
As we discussed, however, $\calR_\omega$ will not be conserved if
the multi-component matter or fields are still dynamical.
It becomes to be conserved when the universe becomes
dominated by a single adiabatic degree of freedom.
Assuming this degree of freedom is encoded in $F$, we have
the equivalence between U$\Omega$S and the comoving slicing,
and hence recover the conformal invariance of $\calR_c$,
namely $\calR_\omega=\calR_c=$constant.

\section{Summary}

To summarize, we have studied the non-perturbative conformal
invariance of the cosmological perturbations. When the universe is
dominated by a single component, the U$\Omega$S and the comoving
slicing coincide and the comoving curvature perturbation $\calR_c$
is conformally invariant fully non-perturbatively.
Consistent with the conformal invariance of $\calR_c$, we have shown
that the $\delta{N}$ formalism gives identical results irrespective
of the choice of conformal frames.

When the universe is dominated by a multiple matter or field
components, the comoving curvature perturbation is no longer conformally
invariant. However, at and after the universe has settled down
to a unique evolutionary trajectory, i.e. 
in an era when there remains only a single adiabatic mode,
we recover the conservation of the comoving curvature perturbation
on super-horizon scales and so it is conformally invariant.

\subsection*{Acknowledgements}

We acknowledge the workshop ``Cosmological Perturbation and
Cosmic Microwave Background'' (YITP-T-10-05) at the Yukawa Institute
for Theoretical Physics, Kyoto University, where the main part of this
work was done,
and the workshop ``WKYC 2011 -- Future of Large Scale Structure Formation'' at Korea Institute for Advanced Study, where this work was initiated and completed.
This work was supported in part by
a Korean-CERN fellowship,
Korea Institute for Advanced Study under the KIAS Scholar program,
the Grant-in-Aid for the Global COE Program at Kyoto University,
``The Next Generation of Physics, Spun from Universality and Emergence''
from the Ministry of Education, Culture, Sports, Science and Technology
(MEXT) of Japan,
JSPS Grant-in-Aid for Scientific Research (A) No.~21244033,
JSPS Grant-in-Aid for Creative Scientific Research No.~19GS0219,
and KRF Grant funded by the Korean Government (KRF-2008-341-C00022).

\appendix
\renewcommand{\theequation}{\thesection.\arabic{equation}}

\section{Conformal decomposition of the spatial metric}
\label{app:confgamma}

Here we give the decomposition of the spatial metric
$\gamma_{ij}$ in a conformally invariant way up to the
 determinant of $\gamma_{ij}$.

As a fiducial background, we take the flat metric,
$\delta_{ij}$. We introduce a traceless matrix $C'_{ij}$,
and decompose it as
\begin{equation}
C'_{ij} = \left( \partial_i\partial_j
 - \frac{1}{3}\delta_{ij}\Delta \right) E + F_{(i,j)}
 + h_{ij} \equiv C_{ij} - \frac{1}{3}\delta_{ij}\Delta{E} \, ,
\end{equation}
where the spatial indices are to be raised or lowered by
the flat metric $\delta_{ij}$, and $F_i$ and $h_{ij}$
satisfy $F^i{}_{,i} = h^i{}_i =h^i{}_{j,i} = 0$.

We write the spatial metric as
\begin{equation}
\gamma_{ij} =a^2(t)e^{2H_L}\widetilde{\gamma}'_{ij}\,,
\end{equation}
where
\begin{equation}
\widetilde{\gamma}'_{ij} = \left( e^{2C'} \right)_{ij}\,.
\end{equation}
Then, we define $\calR$ as the sum of the trace contributions
on the exponent,
\begin{equation}
\label{defcalR}
\calR \equiv H_L - \frac{1}{3}\Delta E \, ,
\end{equation}
and introduce
\begin{equation}
\widetilde{\gamma}_{ij} = \left( e^{2C} \right)_{ij}\,.
\label{defgamma}
\end{equation}

It is then clear that under the conformation transformation,
\begin{equation}
\gamma_{ij}\to\overline{\gamma}_{ij}=\Omega^2\gamma_{ij}\,,
\end{equation}
$\widetilde{\gamma}'_{ij}$ hence $\widetilde{\gamma}_{ij}$
is conformally invariant. In particular, for perturbations
on a given background with $\Omega=\Omega_0e^\omega$,
the conformal transformation affects only $H_L$, or equivalently
$\calR$ as
\begin{equation}
\overline{H}_L=H_L+\omega
\quad
\leftrightarrow
\quad
\overline{\calR}=\calR+\omega\,.
\end{equation}

We can extend the above to the non-flat fiducial metric in a similar way
by covariantizing the derivatives, with the definitions of all
the variables including that of $\calR$, (\ref{defcalR}),
as well as their conformal transformation properties remain the same.

\section{Trajectory in phase space and ${\bm \delta{N}}$}
\label{app:phasespace}

In this appendix, we provide a detailed derivation
of (\ref{deltaN_deltaphideltaphidot}). In many occasions,
it is usually assumed that the number of $e$-folds $N$ is only
a function of $\phi$ and the dependence on $\dot\phi$ is neglected.
But as can be read from (\ref{Rc_deltaphif}) then we lose another
independent degree of freedom in the phase space and cannot find
$\delta\dot\phi$ dependence of $\delta{N}$.

To find the correct dependence of $\delta{N}$ on $\delta\dot\phi$,
we proceed as follows. In the phase space of $\phi$, we consider
that $\phi$ and $\dot\phi$ are functions of two coordinates,
$N$ and $\lambda$. Being the number of $e$-folds, $N$ describes
where we are on a given background trajectory, while $\lambda$ labels
which background trajectory we follow.
Then, formally we can write $\delta\phi$ as
\begin{equation}\label{deltaphi}
\delta\phi = \frac{\partial\phi}{\partial{N}}\delta{N}
 + \frac{\partial\phi}{\partial\lambda}\delta\lambda
 + \frac{1}{2}\frac{\partial^2\phi}{\partial{N}^2}\delta{N}^2
 + \frac{\partial^2\phi}{\partial{N}\partial\lambda}\delta{N}\delta\lambda
 + \frac{1}{2}\frac{\partial^2\phi}{\partial\lambda^2}\delta\lambda^2
 + \cdots \, .
\end{equation}
We can do the same for $\delta\dot\phi$, but using $dN=Hdt$ we have
\begin{equation}
\frac{\delta\dot\phi}{H}
 = \frac{\partial^2\phi}{\partial{N}^2}\delta{N}
 + \frac{\partial^2\phi}{\partial{N}\partial\lambda}\delta\lambda + \cdots \,.
\end{equation}
Multiplying by $\delta{N}$ and using this relation to replace
$\delta{N}\delta\lambda$ term in (\ref{deltaphi}), we have
\begin{equation}\label{deltaphi2}
\delta\phi = \frac{\partial\phi}{\partial{N}}\delta{N} +
\frac{\delta\dot\phi}{H}\delta{N} - \frac{1}{2}
\frac{\partial^2\phi}{\partial{N}^2}\delta{N}^2 + \cdots \, ,
\end{equation}
where we have omitted higher order derivative terms as well as pure
$\delta\lambda$ terms, because now we only consider the perturbation
 {\em along} a given trajectory.

Therefore, from (\ref{deltaphi2}) we can write $\delta{N}$ up to
second order as
\begin{equation}
\frac{\dot\phi_0}{H}
\left( 1 + \frac{\delta\dot\phi}{\dot\phi_0} \right) \delta{N}
 = \delta\phi + \frac{1}{2H} \frac{d}{dt}
 \left( \frac{\dot\phi_0}{H} \right)\delta{N}^2 \, .
\end{equation}
Perturbatively expanding, we can find $\delta{N}$ as
\begin{equation}
-\delta{N} = \frac{H}{\dot\phi_0}\delta\phi
 - \frac{H}{\dot\phi_0^2}\delta\phi\delta\dot\phi
 + \frac{H^2}{2\dot\phi_0^3}
\frac{d}{dt} \left( \frac{\dot\phi_0}{H} \right)\delta\phi^2 + \cdots \,,
\end{equation}
where we have reversed the time order, $\delta{N} \to -\delta{N}$,
because in the context of the $\delta{N}$ formalism it is defined to
be the variation of $N$ due to the variation of the {\em initial}
conditions. Thus we can correctly extract the $\delta\dot\phi$
dependence of $\delta{N}$. Note that we could have found this dependence
by considering $\dot\phi$ as an independent degree of freedom in
the phase space. Once the trajectory follows the attractor,
$\dot\phi_0$ can be written as a function of $\phi_0$ and we may
follow the naive expansion
$\delta{N} = N'\delta\phi + N''\delta\phi^2/2 + \cdots$.

\section{$f(R)$ gravity in Brans-Dicke form}
\label{sec:fR}

It is known that $f(R)$ gravity can be cast into the form of a
Brans-Dicke type theory. Here let us recapitulate it.

We can rewrite the $f(R)$ gravity action, (\ref{eq:fR}),
by introducing an auxiliary scalar field as
\begin{equation}
S=\int d^4x\sqrt{-g}\frac{1}{2}\left[f(s)+\lambda(R-s)\right]\,.
\end{equation}
Then the variation with respect to $\lambda$ gives $R=s$,
and we recover the original action. On the other hand,
if we take the variation with respect to $s$, we obtain
the constraint,
\begin{equation}
\frac{df(s)}{ds}-\lambda=0\,.
\end{equation}
Using this to eliminate $\lambda$ from the above action, we obtain
\begin{equation}
S=\int d^4x\sqrt{-g}\frac{1}{2}\left[\frac{df(s)}{ds}R
+f(s)-s\frac{df(s)}{ds}\right]\,.
\label{eq:fR2}
\end{equation}

If we further introduce a scalar field $\Phi$ by $\Phi\equiv df/ds$,
then the above action is rewritten again as
\begin{equation}
S=\int d^4x\sqrt{-g}\frac{1}{2}\left[\Phi R - I(\Phi)\right]\,,
\label{BDaction}
\end{equation}
where $I(\Phi)\equiv s\,df(s)/ds-f(s)$.
This is a Brans-Dicke scalar-tensor theory with a non-vanishing potential
$V=I(\Phi)/2$, with the Brans-Dicke parameter $w_{\rm BD}=0$.

The field equations are
\begin{align}
& R=\frac{dI(\Phi)}{d\Phi} \, ,
\\
&\Phi\, G_{\mu\nu}-\frac{1}{2}g_{\mu\nu}I(\Phi)
+g_{\mu\nu}\nabla^{\gamma}\nabla_{\gamma}\Phi
-\nabla_{\mu}\nabla_{\nu}\Phi=0\,.
\end{align}
This shows that $\Phi$ is a dynamical field despite the
absence of an apparent kinetic term in (\ref{BDaction}).

We have not performed a conformal transformation so far.
Hence the metric is still the original metric $g_{\mu\nu}$.
Therefore the $f(R)$ action~(\ref{eq:fR}) and
the $F(\Phi)R$ action~(\ref{BDaction}) are completely
equivalent to each other.

With the conformal transformation of $\overline{g}_{\mu\nu}=\Omega^2g_{\mu\nu}$,
with $\Omega^2=\Phi=e^{\sqrt{2/3}\,\psi}$,
the action~(\ref{BDaction}) can be transformed to the Einstein frame,
\begin{equation}
S=\int d^4x \sqrt{-\overline{g}}
\left[\frac{1}{2}\overline{R}
 -\frac{1}{2}\overline{g}^{\mu\nu}\psi_{,\mu}\psi_{,\nu}-\overline{V}(\psi)\right]\,.
\end{equation}

\end{document}